\documentclass[twocolumn,notitlepage,showpacs,aps,longbibliography,prd]{revtex4-1}

\usepackage{epsf}
\usepackage{amsmath}
\usepackage{amsfonts}
\usepackage{amssymb}
\usepackage{bm}
\usepackage{bbm}
\usepackage{nicefrac}
\usepackage{xcolor}
\usepackage{pifont}
\usepackage{graphicx}
\usepackage{enumerate}
\usepackage{dcolumn}
\usepackage{maybemath}
\usepackage{xfrac}
\usepackage{siunitx}
\usepackage{soul}
\usepackage{slashed}

\usepackage{fancyvrb}

\usepackage{hyperref}

\usepackage{upgreek}

\newcommand{\aoc}{{\tilde a}_0}

\newcommand{\DD}{\mathrm{D}}
\newcommand{\NR}{\mathrm{NR}}

\newcommand{\calA}{\mathcal{A}}
\newcommand{\calM}{\mathcal{M}}
\newcommand{\calN}{\mathcal{N}}
\newcommand{\calT}{\mathcal{T}}

\usepackage{color}

\newcommand{\TT}{{\rm T}}
\newcommand{\FF}{{\rm FF}}
\newcommand{\rmR}{{\rm R}}

\newcommand{\xy}{{\parallel}}

\newcommand{\plus}{{\mbox{\scriptsize $(+)$}}}
\newcommand{\minus}{{\mbox{\scriptsize $(-)$}}}

\newcommand{\magnetron}{{\mathrm{m}}}

\newcommand{\hate}{\hat{\mathrm e}}

\newcommand{\dd}{\mathrm{d}}
\newcommand{\ee}{\mathrm{e}}
\newcommand{\ii}{\mathrm{i}}

\newcommand{\qq}{\ell}

\newcommand{\PT}{{\mathrm{T}}}

\newcommand{\calO}{\mathcal{O}}

\definecolor{garrosgreen}{rgb}{0.1, 0.4, 0.1}
\definecolor{dartmouthgreen}{rgb}{0.05, 0.5, 0.06}
\definecolor{duelferred}{rgb}{0.7, 0.2, 0.1}
\definecolor{cambridgeblue}{rgb}{0.1, 0.3, 1.0}
\definecolor{oxfordblue}{rgb}{0.05, 0.2, 0.7}

\definecolor{irishgreen}{rgb}{0.1, 0.7, 0.3}

\begin{document}


\title{Quantum Electrodynamic Corrections to Cyclotron States in a Penning Trap}

\author{Ulrich D. Jentschura}
\affiliation{Department of Physics and LAMOR,
Missouri University of Science and Technology,
Rolla, Missouri 65409, USA}

\author{Christopher Moore} 
\affiliation{Department of Physics and LAMOR,
Missouri University of Science and Technology,
Rolla, Missouri 65409, USA}

\begin{abstract}
We analyze the leading and higher-order 
quantum electrodynamic corrections to the energy levels 
for a single electron bound in a Penning trap,
including the Bethe logarithm correction due to 
virtual excitations of the reference quantum cyclotron state.
The effective coupling parameter $\alpha_c$ 
in the Penning trap is identified as 
the square root of the ratio of the cyclotron frequency,
converted to an energy via multiplication by 
the Planck constant, to the electron rest mass 
energy. We find a large, state-independent, 
logarithmic one-loop self-energy correction of order 
$\alpha \, \alpha_c^4 \, m c^2 \ln(\alpha_c^{-2})$,
where $m$ is the electron rest mass and $c$ is the 
speed of light. Furthermore, we find 
a state-independent ``trapped'' Bethe logarithm.
We also obtain a state-dependent higher-order logarithmic 
self-energy correction of order
$\alpha \, \alpha_c^6 \, m c^2 \, 
\ln(\alpha_c^{-2})$.
In the high-energy part of the bound-state self energy,
we need to consider terms with up to six magnetic 
interaction vertices inside the virtual photon loop.
\end{abstract}

\maketitle

\small

\tableofcontents

\normalsize

%
%
\section{Introduction}
\label{sec1}

Relativistic and quantum electrodynamic
corrections to the quantum cyclotron energy levels
in a Penning trap are of essential importance
for the determination of fundamental 
physical 
constants~\cite{BrGa1982,BrGaHeTa1985,%
BrGa1986,GaEtAl2006everything,HaFoGa2008,FaEtAl2023}.
In a recent article~\cite{WiMoJe2022}, 
higher-order relativistic corrections 
for the energy levels in a quantum cyclotron
have been analyzed.
Here, our goal is to supplement the preceding analysis~\cite{WiMoJe2022}
by a calculation of the intricate 
and notoriously problematic 
quantum electrodynamic (QED) 
corrections to the quantum cyclotron energy levels
inside the Penning trap.
In our calculations, we use expansion parameters
which allow us to initiate a systematic 
classification of the correction terms,
in terms of a semi-analytic expansion 
in terms of a ``trapped fine-structure 
constant $\alpha_c$'', and a
cyclotron scaling parameter $\xi_c$,
as well as an axial
scaling parameter~$\xi_z$.
These parameters replace and supplement
the QED coupling parameter, which is the 
fine-structure constant $\alpha$.

As already anticipated,
the effective coupling parameter in a 
quantum cyclotron could be identified as 
the maximum of the cyclotron ($c$) and the 
axial ($z$) coupling constants.
In particular, one may identify 
the coupling parameters
(in a unit system with $\hbar = c = \epsilon_0 = 1$)
\begin{equation}
\label{defalpha}
\alpha_c = \sqrt{\frac{\omega_c}{m}} \,,
\qquad
\alpha_z = \sqrt{\frac{\omega_z}{m}} \,,
\qquad
\omega_c = \frac{|e| \, B_\TT}{m} \,,
\end{equation}
which depend on the cyclotron frequency $\omega_c$,
and the axial frequency $\omega_z$.
The cyclotron frequency~\cite{BrGa1986}
is $\omega_c = |e| \, B_\PT/m$,
where $e$ is the electron charge,
$|e| = -e$ is the (positive) elementary charge,
$B_\PT$ is the magnetic field in the Penning trap,
and $m$ is the electron mass.

The hierarchy of
typical frequencies in a Penning trap~\cite{BrGa1986,WiMoJe2022}
implies that the magnetron 
frequency $\omega_\magnetron$ is much smaller than the 
axial frequency $\omega_z$.
Following the conventions of Ref.~\cite{BrGa1986},
we define the corrected cyclotron frequency 
as $\omega_\plus$,
we define the corrected magnetron frequency 
as $\omega_\magnetron = \omega_\minus$, where
\begin{subequations}
\begin{align}
\omega_\plus =& \; \frac12 \, \left( \omega_c +
\sqrt{ \omega_c^2 - 2 \omega_z^2 } \right) \,,
\\[0.1133ex]
\label{omega_m}
\omega_\minus =& \; \omega_\magnetron = 
\frac12 \, \left( \omega_c -
\sqrt{ \omega_c^2 - 2 \omega_z^2 } \right) 
\approx \frac{ \omega_z^2 }{ 2 \omega_c } \,.
\end{align}
\end{subequations}
The magnetron frequency is, typically, much smaller than the 
cyclotron frequency~$\omega_c$.
One defines the generalized coupling parameter
\begin{equation}
\alpha_\magnetron = \sqrt{ \frac{\omega_\magnetron}{m} } 
\end{equation}
for the magnetron frequency.
We assume the following hierarchy
to be fulfilled (see Ref.~\cite{BrGa1986}),
\begin{equation}
\alpha_\magnetron \ll \alpha_z \ll \alpha_c \,.
\end{equation}
We also define scaling parameters $\xi_z$ and $\xi_\magnetron$ 
by
\begin{align}
\alpha_z =& \; \xi_z \, \alpha_c \,,
\qquad
\alpha_\magnetron = \xi_\magnetron \, \alpha_c \,,
\\
\xi_\magnetron =& \; \frac{1}{\sqrt{2}}
\left( 1 - \sqrt{ 1 - 2 \xi_z^4 } \right)^{1/2} \approx
\frac{ \xi_z^2 }{ \sqrt{2} } \,.
\end{align}
The hierarchy of the frequencies allows us to perform
a systematic expansion in terms of $\alpha_c$,
$\xi_z$ and $\xi_\magnetron$,
as well as the coupling parameter of quantum electrodynamics,
that is, the fine-structure constant $\alpha$.
The expansion in $\xi_\magnetron$ 
gives rise to an expansion in $\xi_z^2$,
and we can henceforth use the parameter $\xi_z$
in order to universally describe the parametrically suppressed
effects due to both the axial as well as magnetron motions.

A remark is in order, concerning the 
anticipated results of our studies.
It is well known~\cite{Be1947,Pa1993,JePa1996} that the leading 
logarithmic quantum electrodynamic self-energy 
correction to hydrogen energy levels is
proportional to 
(in natural units, $\hbar = c = \epsilon_0 = 1$, which are used here)
\begin{equation}
\Delta E_{\rm QED} \sim \alpha \, (Z \alpha)^4 \, m \, 
\ln[(Z \alpha)^{-2}] \,,
\end{equation}
where $\alpha$ is the fine-structure constant,
and $Z$ is the nuclear charge number.
We here anticipate that we shall find the 
following, analogous scaling for the 
leading quantum electrodynamic self-energy correction 
to quantum cyclotron levels in a Penning trap,
\begin{equation}
\Delta E_{\rm QED} \sim \alpha \, \alpha_c^4 \, m \, \ln(\alpha_c^{-2}) \,,
\end{equation}
where the coefficient $\alpha$ is due to the 
absorption and emission of the virtual 
photon, and the factors of $\alpha_c$ describe the 
binding to the trap fields,
which is typically smaller than the 
coupling parameter $\alpha$ for atoms.
It is our goal to calculate these energy shifts.

This paper is organized as follows.
In Sec.~\ref{sec2}, we present a brief 
review of the quantum cyclotron states which enter
our formalism.
Vacuum-polarization corrections are negligible
for quantum cyclotron states, for reasons outlined 
in Sec.~\ref{sec3}.
Self-energy effects are discussed in Sec.~\ref{sec4};
these constitute the dominant radiative 
corrections for quantum cyclotron states.
Conclusions are reserved for Sec.~\ref{sec5}.

%
%
\section{Quantum Cyclotron Levels}
\label{sec2}

In order to understand the quantum cyclotron levels
inside a Penning trap, it is, first of all, necessary to 
remember that the kinetic momentum is given by
\begin{equation}
\label{piPTdef}
\vec\pi_\PT = \vec p - e \vec A_\PT
= \vec p - \frac{e}{2} \, ( \vec B_\PT \times \vec r ) \,,
\end{equation}
where $\vec A_\PT = \tfrac12 (\vec B_\PT \times \vec r)$ 
is the vector potential,
$\vec B_\PT = B_\PT \, \hate_z$ is the magnetic field
in the trap, and $\vec p = -\ii \vec\nabla$ is the kinetic 
momentum operator. The kinetic momentum $\vec \pi_\PT$ enters
the interaction Hamiltonian 
describing the coupling of the bound electron (inside the 
Penning trap) to the quantized electromagnetic field.

The quadrupole electric field in the trap 
is attractive along the $z$ axis 
and repulsive in the $xy$ plane,
\begin{subequations}
\label{potV}
\begin{align}
V =& \; = V_z + V_\xy \,,
\qquad
\vec\nabla^2 V = 0 \,,
\\[0.1133ex]
\label{Vpar}
V_z =& \; \frac12 m \omega_z^2 z^2
\qquad
V_\xy = - \frac14 m \omega_z^2 \rho^2 \,.
\end{align}
\end{subequations}
The unperturbed Hamiltonian is given as follows,
\begin{equation}
\label{H0rep1}
H_0 = \frac{( \vec\sigma \cdot \vec \pi_\PT)^2 }{2 m} + V
- \frac{e}{2m} \, \kappa \, \vec\sigma\cdot \vec B_\PT \,.
\end{equation}
Eigenstates of the unperturbed Hamiltonian
$H_0$ are described~\cite{BrGa1986} by 
four quantum numbers: 
the axial quantum number $k$, the magnetron quantum number $\qq$,
the cyclotron quantum number $n$, and
the spin projection quantum number $s = \pm 1$.
These take on the following values:
$k = 0,1,2,\dots$ (axial),
$\qq = 0,1,2,\dots$ (magnetron),
$n = 0,1,2,\dots$ (cyclotron),
and $s = \pm 1$ (spin).
We recall, from Ref.~\cite{WiMoJe2022}, 
the energy eigenvalues of $H_0$,
\begin{align}
\label{Eklns}
E_{k \ell n s} =& \; \omega_c (1+\kappa) \, \frac{s}{2}
+ \omega_\plus \left( n + \frac12 \right) 
\nonumber\\[0.1133ex]
& \; + \omega_z \left( k + \frac12 \right)
- \omega_\minus \left( \qq + \frac12 \right) \,.
\end{align}
It is of note that, in view of the repulsive character of the 
quadrupole potential,
these eigenvalues  are not bounded from below.
We use the conventions of Refs.~\cite{BrGa1986,WiMoJe2022},
for the cyclotron lowering and raising operators 
$a_\plus$ and $a^\dagger_\plus$,
the axial lowering and raising operators 
$a_z$ and $a^\dagger_z$, and the 
magnetron lowering and raising operators 
$a_\minus$ and $a^\dagger_\minus$.
The eigenstates of the unperturbed Hamiltonian 
are given as follows,
\begin{align}
\label{psidef}
\psi_{k \ell n s}(\vec r) = & \;
\frac{\left( a^\dagger_\plus \right)^n}{\sqrt{n!}} \,
\frac{\left( a^\dagger_z \right)^k}{\sqrt{k!}} \,
\frac{\left( a^\dagger_\minus \right)^q}{\sqrt{q!}} \,
\psi_0(\vec r) \, \chi_{s/2} \,,
\nonumber\\[0.1133ex]
\chi_{1/2} =& \; \left( \begin{array}{c} 1 \\ 0 \end{array} \right) \,,
\qquad
\chi_{-1/2} = \left( \begin{array}{c} 0 \\ 1 \end{array} \right) \,.
\end{align}
The orbital part of the ground-state wave function is 
\begin{align}
\psi_0(\vec r) =& \;
\sqrt{ \frac{ m \sqrt{ \omega_c^2 - 2 \omega_z^2 } }{ 2 \pi } } \,
\exp\left( - \frac{m}{4} \sqrt{ \omega_c^2 - 2 \omega_z^2 } \, \rho^2 \right) 
\nonumber\\[0.1133ex]
& \; \times \left( \frac{ m \omega_z }{\pi} \right)^{1/4} \,
\exp\left( - \frac12 m \omega_z z^2 \right) \,.
\end{align}
The spin-up sublevel of the $n$th cyclotron ground state,
and the spin-down sublevel of the 
$(n+1)$st excited cyclotron state,
are quasi-degenerate and 
of interest for spectroscopy and determination
of the anomalous magnetic moment of the 
electron~\cite{BrGa1982,BrGaHeTa1985,GaEtAl2006everything,HaFoGa2008,FaEtAl2023}.

%
%

\section{Vacuum Polarization}
\label{sec3}

For atomic bound states,
quantum electrodynamic energy shifts
are naturally separated into vacuum-polarization
and self-energy corrections. 
The vacuum-polarization shift of a hydrogenic
energy level is due to the screening of the 
proton's charge by virtual electron-positron
pairs. The closer the electron is to the nucleus,
the less pronounced is the screening of the bare proton charge,
and the stronger is the (corrected) Coulomb potential.
The dominant contribution to the one-loop 
effect is described by the Uehling potential~\cite{Ue1935}.
In a Penning trap, the potential is generated by 
the trap electrodes in addition to the 
axial magnetic field.
Hence, the electron, on its quantum cyclotron
orbit, is always sufficiently far away from any 
other charged particle that the vacuum-polarization 
energy shift can be safely neglected.
This statement can be quantified as follows.

The long-range tail of the Uehling potential
is given as follows~\cite{Je2011aop1},
\begin{equation}
V_U(r) \approx 
-\frac{\alpha (Z \alpha) m}{4 \sqrt{\pi}} \,
\frac{\exp(- m r)}{(m \, r)^{5/2}} \,,
\qquad r \to \infty \,,
\end{equation}
where $r$ is the distance from the nucleus.
The one-loop Uehling correction needs to be  
compared to the Coulomb potential,
\begin{equation}
V_C(r) \approx -\frac{Z \alpha m}{( m r) } \,,
\end{equation}
leading to the relative correction,
\begin{equation}
\frac{V_U(r)}{V_C(r)} \approx 
\frac{\alpha}{ 4 \sqrt{\pi} \, (m r)^{3/2} } \,
\exp(- m r) \,,
\qquad r \to \infty \,,
\end{equation}
A typical Penning 
trap dimension~\cite{BrGa1986}
is of the order of about $\langle r \rangle \sim 1 \, {\rm cm}$,
while the quantity $m \, r$ is dimensionless
in natural units. When converted to Syst\`{e}me International
(mksA) units, one realizes that $m$ takes the role 
of the inverse of the reduced Compton wavelength 
of the electron,

\begin{equation}
m r = \frac{r}{\lambdabar_e} = 
2.59 \times 10^{12} \, \times R \,,
\qquad 
R = \frac{r}{1 \, {\rm m}} \,,
\end{equation}
where $R$ is measured in meters. 
For $R$ being on the order of $1 \, {\rm cm}$,
one has $m r$ on the order of $10^{10}$.
The quantity 
\begin{equation}
\exp(- m \langle r \rangle) \sim
\exp(- 10^{10} ) \approx
10^{-4.3 \times 10^{9}} 
\end{equation}
is very small indeed. Its smallness illustrates that,
because of the exponential expression of the one-loop
vacuum-polarization correction to the quadrupole trap potential,
the vacuum-polarization corrections can be neglected.
The same exponential suppression factor $\exp(-mr )$
enters the magnetic photon exchange~\cite{Je2011pra}
which is the basis for the magnetic field of the 
trap. Therefore, vacuum-polarization corrections
can be safely neglected for quantum cyclotron levels.

%
%
\section{Self Energy}
\label{sec4}

\subsection{Orientation}

Inspired by the formalism pertinent
to bound states in a Coulomb field~\cite{JePa1996,JeSoMo1997},
we write the semi-analytic expansion of the 
one-loop bound-state energy shift 
of a quantum cyclotron state
as follows,
\begin{equation}
\label{ESEcoeff}
E_{\rm SE} = \frac{\alpha}{\pi} \, m
\sum_{rs} \calA_{rs} \, ( \alpha_c )^r \,
\ln^s\left(\alpha_c^{-2} \right)
\end{equation}
where the first subscript of the $\calA$ coefficients
counts the power of $\alpha_c$, and the second subscript
indicates the power of the logarithm 
$\ln\left(\alpha_c^{-2} \right)$.

The $\calA_{rs}$ coefficients are analogous to the 
coefficients $A_{rs}$ used in Lamb shift calculations
for hydrogenlike systems (see Sec.~15.4 of Ref.~\cite{JeAd2022book}).
For the electron in the Penning trap,
the role of the Coulombic coupling parameter $Z \alpha$ 
is taken by the cyclotron coupling parameter $\alpha_c$.
In Lamb shift calculations for hydrogenic systems,
one scales out a factor $1/n^3$ from the coefficients,
where $n$ is the principal quantum number.
This reflects on the typical scaling of
quantum electrodynamic energy corrections
in hydrogenlike systems.
In the Penning trap, the role of $n$ is
played by the cyclotron quantum number.
However, there is a decisive 
difference: For the Penning trap,
no $1/n^3$ dependence is incurred, and in fact,
some logarithmic coefficients are seen to increase
with $n$, not decrease as is typically 
the case in Coulombic bound systems. 
We thus do not scale out $1/n^3$ in the definition
of the $\calA_{rs}$ coefficients.

The leading self-energy coefficient is 
seen to be $\calA_{20}$, and is due to 
the leading Schwinger term~\cite{Sc1948}
in the anomalous magnetic moment of the electron.
Here, we focus on the coefficients
$\calA_{20}$, $\calA_{41}$, $\calA_{40}$, and $\calA_{61}$,
which constitute the 
leading nonvanishing coefficients 
for a general quantum cyclotron state.
The higher-order nonlogarithmic coefficients possess
an expansion in powers of $\xi_z$,
e.g.,
$\calA_{40} = \left. \calA_{40}  \right|_{\xi_z=0} 
+ \calO(\xi_z)$.
We here evaluate $\calA_{20}$, $\calA_{41}$ and $\calA_{61}$, 
and $\calA_{40}$ in the leading order in $\xi_z$,
and partial results for the corrections 
proportional to $\xi_\magnetron$ and $\xi_z$.
The Bethe logarithm inside the Penning trap
is seen to contribute to $\calA_{40}$,
albeit only at order $\xi_z$. 
Indeed, in the leading order in the expansion
in $\xi_z$, the Bethe logarithm in the Penning
trap will be seen to vanish.
Our result for the Bethe logarithm is numerically small and,
somewhat surprisingly, 
state-independent. The contribution of the Bethe logarithm
is thus not visible in any transitions
among quantum cyclotron states.
Let us anticipate some results which will be
derived in the following, in order to 
lay out the work program of our article.
Indeed, we obtain two contributions
to the order-$\xi_z$ correction to $\calA_{40}$,
one from a higher-order anomalous magnetic 
moment term, and a second one from the 
Bethe logarithm. There might be another 
contribution to the order-$\xi_z$ correction to $\calA_{40}$,
from the high-energy part, which we evaluate
only to leading order in $\xi_z$. The evaluation of the 
complete order-$\xi_z$ correction to $\calA_{40}$ 
will be left for a future investigation.
The dominant state-dependent correction, 
in the leading order in $\xi_\magnetron$ and $\xi_z$,
is found to be given by the $\calA_{61}$ coefficient.

The coefficients $\calA_{20}$, $\calA_{41}$, 
$\calA_{40}$, and $\calA_{61}$, determined here,
constitute the leading nonvanishing coefficients
for the self-energy effect.
Coefficients of odd order in $\alpha_c$
such as $\calA_{31}$ and $\calA_{30}$,
as well as $\calA_{50}$, vanish. 
[By odd order, in general, we refer to an 
odd integer $r$ in Eq.~\eqref{ESEcoeff}.]
A brief discussion on this point,
mostly based on the high-energy part discussed 
in Sec.~\ref{sec4C}, is in order.
The $\calA_{20}$ coefficients is a consequence of the 
Schwinger term~\cite{Sc1948} which enters
at lower order because the main binding potential
involves the magnetic trap field.
Operators in the high-energy part
can be expanded in the (vector-)potential
$\Gamma_\PT$ defined in Eq.~\eqref{GammaPT}
and in the momentum operators.
For quantum cyclotron states, momentum 
and potential operators (coordinates)
can be expressed in terms of raising and 
lowering operators of the cyclotron and 
magnetron quantum numbers~\cite{WiMoJe2022},
and hence, all of these matrix elements 
are convergent (in arbitrarily high orders 
in $\alpha_c$). Because of symmetry reasons,
matrix elements which would otherwise
lead to an odd power of $\alpha_c$ vanish.
An example would be terms of third order in 
the momentum operators, whose matrix 
element on the reference state vanishes due to parity.
[Odd orders in $\alpha_c$ would otherwise
correspond to half-integer powers in 
$\omega_c$, in view of Eq.~\eqref{defalpha}.]

The terms $\calA_{41}$ and $\calA_{40}$ 
are generated by mechanism much in 
analogy to those at work 
in Coulombic bound systems (see Chaps.~4 and~11 of 
Ref.~\cite{JeAd2022book}).
Finally, one might ask why the term $\calA_{50}$
vanishes for quantum cyclotron levels
in a Penning trap, while the corresponding 
term $A_{50}$ for Coulombic systems
is nonvanishing (see Chap.~15 of 
Ref.~\cite{JeAd2022book} and Ref.~\cite{BaBeFe1953}).
A closer inspection reveals that the emergence of 
the $A_{50}$ term (for radially symmetric $S$ states in 
Coulombic systems) is caused by the singularity 
of the Coulomb potential and of the hydrogen eigenstates.
The singularity of the Coulomb potential
eventually leads to the divergence of matrix elements
$\langle \vec p^{\,6} \rangle$ when evaluated on
reference $S$ states, which prevents the 
direct expansion of the high-energy part of the 
self-energy (Sec.~\ref{sec4C}) in powers of momentum operators beyond 
fourth order. For quantum cyclotron states, however, the 
potential has no singularity at the origin,
and hence, matrix elements of arbitrarily 
high orders in the momenta are convergent.
No term of fifth order 
in $\alpha_c$ is generated ($\calA_{50}=0$).  

%
%
\subsection{Form Factor Treatment}

In typical cases,
the self-energy shift of a bound electronic state
is the sum of a high-energy part 
(due to virtual photons of high energy), 
and a low-energy part
(due to virtual photons whose energy is of the 
same order as the quantum cyclotron binding energy).
The matching of the high- and low-energy parts
is quite problematic (see footnote~13 on p.~777 
of Ref.~\cite{Fe1949}).
One may complete the matching based 
on photon mass or photon energy regulations,
or in dimensional regularization (see Chaps.~4 and 
11 of Ref.~\cite{JeAd2022book}).
In many cases, the high-energy 
part can be handled on the basis 
of a form-factor approach
[see, e.g., Eq.~(3) of Ref.~\cite{JePa2002}],
provided the photon mass and photon
energy cutoffs are properly matched
[see, e.g., Eqs.~(32) and~(33) of Ref.~\cite{JePa2002}].

In the case of a Penning trap, one needs
to reformulate the effective Dirac Hamiltonian
obtained from a form-factor treatment,
because there is both a nonvanishing vector potential,
as well as an electric quadrupole potential,
present in the trap.
Let us discuss in some detail.
We start with the structure of the 
electromagnetic field-strength tensor, 
in a component-wise representation, 
\begin{equation}
\label{Fmunucomp_contra}
F^{\mu\nu} =
[ \left( \begin{array}{cccc}
0 & -E^x & -E^y & -E^z \\
E^x & 0 & -c B^z & c B^y \\
E^y & c B^z & 0 & -c B^x \\
E^z & -c B^y & c B^x & 0
\end{array} \right) ]^{\mu\nu} \,.
\end{equation}
For the Dirac matrices, we use the Dirac representation, where
\begin{equation}
\label{dirac}
\gamma^0 = \beta = \left( \begin{array}{cc}
\mathbbm{1}_{2 \times 2} & 0 \\
0 & -\mathbbm{1}_{2 \times 2} 
\end{array} \right) \,,
\qquad
\gamma^i = \left( \begin{array}{cc}
0 & \sigma^i \\
-\sigma^i & 0 \\
\end{array} \right) \,.
\end{equation}
The $\sigma^i$ are the Pauli matrices,
and Latin indices are spatial ($i=1,2,3$).
The spin matrices are defined as $\sigma^{\mu\nu} =
\tfrac{\ii}{2} [ \gamma^\mu, \gamma^\nu ]$,
and the Dirac $\alpha$ and $\Sigma$ matrices are
\begin{equation}
\alpha^i = \left( \begin{array}{cc}
0 & \sigma^i \\
\sigma^i & 0 \\
\end{array} \right) \,,
\qquad
\Sigma^i = \left( \begin{array}{cc}
\sigma^i & 0 \\
0 & \sigma^i \\
\end{array} \right) \,,
\end{equation}
One derives the relation
\begin{equation}
\sigma_{\mu \nu} F^{\mu \nu} = 
2 \ii \, \vec\alpha \cdot \vec E- 
2 \vec\Sigma\cdot \vec B \,.
\end{equation}
The replacement for the $\gamma^\mu$ matrix 
at the vertex (Greek indices are spatio-temporal, $\mu  = 0,1,2,3$) is
(see Chap.~10 of Ref.~\cite{JeAd2022book})
\begin{equation}
\gamma^\mu \to \gamma^\mu \, F_1(q^2)
+ \ii \frac{\sigma^{\mu\nu}}{2 m} q_\nu \, F_2(q^2) \,.
\end{equation}
Here, $F_1$ is the Dirac form factor, while
$F_2$ is the Pauli form factor.
In coordinate space, the interaction Hamiltonian is obtained 
from the replacement $q^2 \to -\vec q^{\,2} \to \vec\nabla^2$,
$q_\nu \to \ii \partial_\nu$, and results in
\begin{multline}
e \gamma^\mu A_\mu \to 
F_1(\vec\nabla^2) \, e \, \gamma^\mu \, A_\mu(\vec r)
\\
+ F_2(\vec\nabla^2) \, \frac{e}{2 m} \,
\left( \ii \, \vec\alpha \cdot \vec E-
\vec\Sigma\cdot \vec B \right) \,.
\end{multline}
The interaction Hamiltonian of quantum 
electrodynamics is $j^\mu \, A_\mu = 
e \, \psi^\dagger \, \gamma^0 \, \gamma^\mu \, A_\mu \, \psi$.
So, the contribution to the Hamiltonian, 
in the space of the scalar product equipped with 
$\psi$ and $\psi^\dagger$, is obtained 
from the expression $e \, \gamma^\mu \, A_\mu$,
via multiplication by $\gamma^0$.
Hence, the modified Dirac Hamiltonian reads as
\begin{align}
\label{HDmCoulomb}
H_{\rmR} =& \; \vec{\alpha} \cdot \vec{p}
+ \beta\,m + F_1(\vec{\nabla}^2) \, W 
\nonumber\\[0.1133ex]
& \; + F_2(\vec{\nabla}^2) \, \frac{e}{2\,m} \,
\left( {\mathrm i}\, \vec{\gamma} \cdot \vec{E}
- \beta \, \vec\sigma \cdot \vec B \right) \,,
\nonumber\\[0.1133ex]
W =& \; e \, \gamma^0 \, \gamma^\mu \, A_\mu
= e \, \alpha^\mu \, A_\mu \,,
\end{align}
We now consider a vector potential
$A^\mu = (A^0(\vec r), \vec A(\vec r))$, 
where $A^0(\vec r) = \Phi(\vec r)$ is 
the quadrupole potential of the Penning trap and 
$\vec A(\vec r) = \vec A_\TT(\vec r) = 
\tfrac12 (\vec B_\TT \times \vec r)$
is the vector potential corresponding to the 
magnetic field of the trap.
One can rewrite the radiatively corrected
Hamiltonian as
\begin{align}
\label{HDmFIELD}
H_{\rmR} =& \; \vec{\alpha} \cdot
\left[\vec{p} - e \, F_1(\vec{\nabla}^2) \, \vec{A}_\TT(\vec r) \right]
+ \beta\,m + F_1(\vec{\nabla}^2) \, e A^0_\TT(\vec r)
\nonumber\\[0.1133ex]
& \; + F_2(\vec{\nabla}^2) \, \frac{e}{2\,m} \,
\left[ \ii \, \vec{\gamma} \cdot \vec{E}(\vec r) -
\beta \, \vec{\Sigma} \cdot \vec{B}(\vec r) \right] \,.
\end{align}
This expression can alternatively be written as
the sum of a covariantly coupled tree-level Hamiltonian
$H_\TT$ and a form-factor correction $H_\FF$,
\begin{subequations}
\begin{align}
H_{\rmR} =& \; H_\TT + H_\FF \,,
\nonumber\\[0.1133ex]
H_\TT = & \; \vec{\alpha} \cdot \vec \pi
+ \beta\,m + e A^0_\TT(\vec r) \,,
\nonumber\\[0.1133ex]
H_{\rm FF} =& \;
[F_1(\vec{\nabla}^2) - 1] \, e A^0_\TT(\vec r) 
- [F_1(\vec{\nabla}^2) - 1] \, 
e \, \vec\alpha \cdot \vec{A}_\TT(\vec r) 
\nonumber\\[0.1133ex]
& \; + F_2(\vec{\nabla}^2) \, \frac{e}{2\,m} \,
\left[ \ii \, \vec{\gamma} \cdot \vec{E}(\vec r) -
\beta \, \vec{\Sigma} \cdot \vec{B}(\vec r) \right] \,.
\end{align}
\end{subequations}
For the nonrelativistic momenta typical of an
electron in a Penning trap, one can 
expand the Dirac form factor $F_1(\vec\nabla^{\,2})$ 
in terms of its argument.
The quadrupole potential of the trap is,
according to Eq.~\eqref{potV},
\begin{equation}
V = e A^0_\TT(\vec r) 
= \frac12 \, m \omega_z^2 \left[z^2 - \frac12 (x^2 + y^2) \right] \,,
\quad
\vec\nabla^2 V = 0 \,.
\end{equation}
So, we can replace
\begin{equation}
[F_1(\vec{\nabla}^2) - 1] \, e A_\TT^0(\vec r) = 0 \,,
\end{equation}
by expansion of the form factor in powers of its argument.
Also, one has
\begin{equation}
\vec A_\TT(\vec r) = \frac12 \, (\vec B_\TT \times \vec r) \,,
\qquad
\vec\nabla^2 \, \vec A_\TT(\vec r) = 0 \,.
\end{equation}
Hence, corrections induced 
by the Dirac form factor vanish for a 
particle bound into a Penning trap.

The only contribution which can 
be evaluated based on the form-factor treatment
concerns the contribution of the anomalous magnetic moment
of the electron to the self energy.
It can be evaluated 
based on a Foldy--Wouthuysen transformation~\cite{WiMoJe2022}
of the radiatively corrected Dirac Hamiltonian 
given in Eq.~(11.40) of Ref.~\cite{JeAd2022book}.
One starts from Eq.~\eqref{HDmFIELD},
approximates~\cite{Sc1949}
\begin{equation}
F_2(0) \approx \kappa = \alpha/(2 \pi) \,,
\end{equation}
and performs a number of unitary transformations
in order to disentangle the particle degrees
of freedom from the antiparticle degrees of freedom.
After the Foldy--Wouthuysen transformation,
one gets two contributions to the Hamiltonian
which are proportional to the 
electron anomalous magnetic moment.
The relevant terms from 
Eqs.~(82) and (87) of Ref.~\cite{WiMoJe2022}
can be summarized in
the radiatively corrected 
anomalous-magnetic moment Hamiltonian $H_\kappa$, 
\begin{multline}
\label{Hkappa}
H_\kappa =  -\frac{e \kappa}{2 m} \vec \sigma \cdot \vec B_\TT
+ \frac{\kappa}{2 m^2} \,
\vec\sigma \cdot ( \vec \nabla V \times \vec\pi_\PT ) 
\\
+ \frac{e \kappa}{4 m^3} \, ( \vec\sigma \cdot \vec\pi_\PT ) \,
( \vec B_\PT \cdot \vec\pi_\PT ) \,.
\end{multline}
In view of the occurrence of the 
scalar product $\vec B_\PT \cdot \vec\pi_\PT = B_\TT p_z$,
the expectation value of the effective
Hamiltonian $H_\kappa$ contains terms which are
linear and quadratic in the 
axial frequency $\omega_z$.
The energy perturbation is obtained as

\begin{equation}
\langle \psi_{k \ell n s} | H_\kappa | \psi_{k \ell n s} \rangle = 
E_{\rm HEP}^{(0)} + E_{\rm HEP}^{(1)} \,,
\end{equation}
where 
\begin{equation}
\label{E0}
E_{\rm HEP}^{(0)} = 2 \kappa s \omega_c
\end{equation}
is the leading term due to the anomalous magnetic moment,
and
\begin{multline}
\label{E1HEP}
E_{\rm HEP}^{(1)} = 
- \frac{s \kappa \omega_c \omega_z}{4 m} (k + \tfrac12) 
\\
-\frac{\omega_z^2 s \kappa}{2 m}
\frac{ \omega_\plus (n + \tfrac12) + \omega_\minus (\ell + \tfrac12) }%
{\omega_\plus - \omega_\minus} \,.
\end{multline}
The two terms after the equal sign
are proportional to $\xi_z^2$ and $\xi_z^4$, respectively.

%
%
\subsection{High--Energy Part}
\label{sec4C}

From Lamb shift calculations for hydrogenic 
bound states~\cite{JePa1996,JeSoMo1997},
we know that in typical self-energy calculations,
the low-energy part, which involves the 
Bethe logarithm, has an ultraviolet
divergence. This ultraviolet divergence 
is compensated by an infrared
divergence of the high-energy part.
Furthermore, for the one-loop self-energy of a hydrogenic bound
state, the infrared divergence of the high-energy 
part can be obtained on the basis 
of an effective potential proportional to 
the infrared slope of the Dirac form factor
(see Chaps.~4 and 11 of Ref.~\cite{JeAd2022book}).

However, we have shown that the Dirac form-factor induced
one-loop correction to the energy 
of a quantum cyclotron state vanishes.
This leaves the question of the correct treatment
of the high-energy part of the 
bound-electron self energy in the 
quantum cyclotron state.

From bound-state calculations for an electron
in a Coulomb field~\cite{JePa1996}, we know that an appropriate treatment 
of the high-energy part consists in the expansion of the one-loop
self-energy operator in terms of the binding 
field. 

In the Feynman gauge, the bound-electron 
self-energy for a quantum cyclotron 
state can be written as
[see Eq.~(15.17) of Ref.~\cite{JeAd2022book}]
\begin{align}
\label{feynman}
\Delta E_{\rm SE} =& \; e^2
\int_{C_F} \frac{\dd^4 k}{(2 \pi)^4 \, \ii} \,
\frac{e^2 \, g_{\mu\nu}}{k^2} \,
\left< \overline\Psi \left| \gamma^\mu \,
\frac{1}{\slashed{\pi} - \slashed{k} - m} \,
\gamma^\nu \right| \Psi \right> 
\nonumber\\[0.1133ex]
& \; - \left< \overline\Psi \left| \delta m \right| \Psi \right> \,.
\end{align}
Here, $C_F$ specifies the Feynman integration 
contour for the photon energy integration. 
The metric is $g_{\mu\nu} = \mathrm{diag}(1,-1,-1,-1)$.
The Dirac matrices are used in the 
Dirac representation given in Eq.~\eqref{dirac}.
The kinetic-momentum four-vector is 
\begin{equation}
\pi^\mu = (E, \vec \pi) \,,
\qquad
\vec\pi = \vec \pi_\TT = \vec p - e \, \vec A_\TT \,,
\end{equation}
where $\vec \pi_\TT$ is defined in 
Eq.~\eqref{piPTdef}.
In the high-energy part, one can expand
the Feynman propagator in powers
of the binding vector potential,
\begin{multline}
\label{nvertex}
\frac{1}{\slashed{\pi} - \slashed{k} - m} =
\frac{1}{\slashed{p} - \slashed{k} - m} 
+ \frac{1}{\slashed{p} - \slashed{k} - m} \,
\Gamma_\TT
\frac{1}{\slashed{p} - \slashed{k} - m} 
\\
+ \frac{1}{\slashed{p} - \slashed{k} - m} \,
\Gamma_\TT \,
\frac{1}{\slashed{p} - \slashed{k} - m} \,
\Gamma_\TT \,
\frac{1}{\slashed{p} - \slashed{k} - m} + 
\\
+ \sum_{n=3}^\infty 
\frac{1}{\slashed{p} - \slashed{k} - m} \,
\left( \Gamma_\TT \,
\frac{1}{\slashed{p} - \slashed{k} - m} \right)^n  \,,
\end{multline}
where
\begin{equation}
\label{GammaPT}
p^\mu = (E, \vec p) \,,
\qquad
\Gamma_\TT = -e \vec \gamma \cdot \vec A_\TT 
\end{equation}
is the Feynman slash of the vector potential 
of the trap. 

\begin{figure}[t!]
\begin{center}
\begin{minipage}{1.0\linewidth}
\begin{center}
\includegraphics[width=0.87\linewidth]{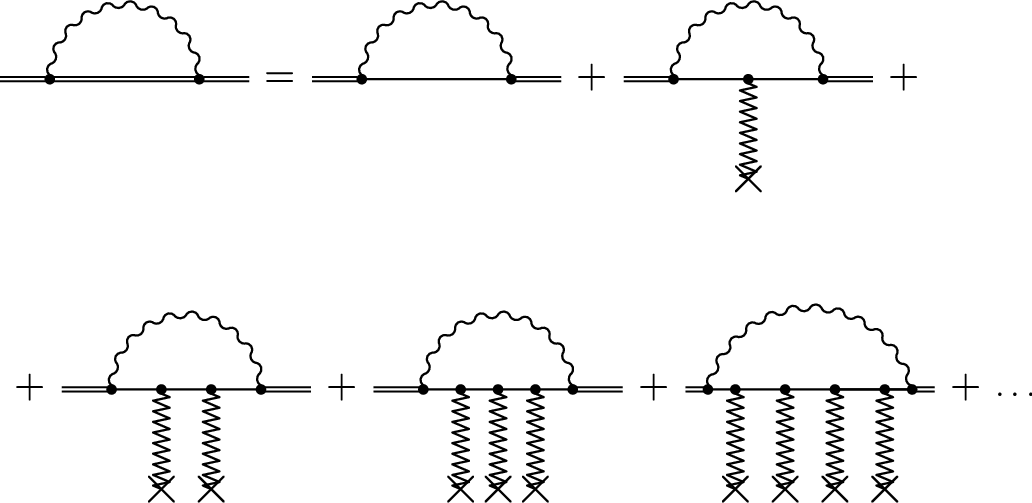}
\caption{\label{fig1}
The figure illustrates the Feynman diagrams 
contributing to the high-energy part 
of the bound-electron self energy 
of a quantum cyclotron state
in magnetic interaction vertices.
These correspond to ascending powers of the 
operator $\Gamma_\TT$ 
as delineated in Eq.~\eqref{nvertex}.}
\end{center}
\end{minipage}
\end{center}
\end{figure}

The mass counter term is $\delta m$, and the
Dirac adjoint is $\overline \psi = \psi^\dagger \, \gamma^0$.
Alternatively, the use of the Feynman contour
can be enforced by the replacements
\begin{equation}
\frac{g_{\mu\nu}}{k^2} \to
\frac{g_{\mu\nu}}{k^2+ \ii\epsilon} \,,
\qquad
\frac{1}{\slashed{\pi} - \slashed{k} - m} \to
\frac{1}{\slashed{\pi} - \slashed{k} - m + \ii \epsilon} \,,
\end{equation}
in the photon and electron propagators.
We use the noncovariant photon
energy cutoff $\epsilon$ introduced in
Refs.~\cite{Pa1993,JePa1996,JeSoMo1997}
which cuts off the Feynman contour
for the photon energy integration at an
infrared cutoff $\epsilon$ which is
of order of the binding energy of the bound
state. The dependence on $\epsilon$ disappears
when the high- and low-energy parts are added.
A further difficulty arises:
For hydrogenic states, the 
operator $\Gamma_\TT$ is replaced by 
$\Gamma_C = +e \gamma^0 \vec A_0 = \gamma^0 (-Z\alpha/r)$,
where $Z$ is the nuclear charge number,
$\alpha$ is the fine-structure constant,
and $r$ is the electron-nucleus distance~\cite{JePa1996,JeSoMo1997}.
One notes that $\Gamma_\TT$ is an odd operator
(connecting upper and lower components of the 
Dirac bispinor), while $\Gamma_C$ 
is an even operator in the 
bispinor basis [see Eq.~\eqref{dirac}].
We must now go into detail and reflect on the 
$Z\alpha$-expansion.
For hydrogenic bound states, the Coulomb potential
scales as $(Z \alpha)^2 \, m$, because 
$r \sim a_0/Z = 1/(Z\alpha m)$, 
where $a_0$ is the Bohr radius. Hence,
every insertion of a power of $\Gamma_C$ 
into the diagram adds two powers of $Z\alpha$.
For the quantum cyclotron state, the expansion
works differently: The role of the coupling parameter
$Z\alpha$ is taken over by $\alpha_c$,
defined in Eq.~\eqref{defalpha}.
One has the following order-of-magnitude estimates:
$B_\TT \sim \alpha_c^2 m^2$, $r \sim 1/(\alpha_c m)$, 
and $|\vec\pi_\TT| \sim \alpha_c \, m$.
However, the matrix 
element $\langle \overline\Psi | \Gamma_\TT 
| \Psi \rangle$ is of order $\alpha_c^2 \, m$ 
and thus, of order of the bound-state 
energy in the quantum cyclotron,
because it connects upper and lower 
components of the Dirac bispinor solution~\cite{Je2023mag1}
(lower components are suppressed by a factor of $\alpha_c$).
Now, while in one occurrence of the 
operator $\Gamma_\TT$, one connects
upper and lower components, 
two such operators connect upper to upper, 
and lower to lower components,
eliminating two powers of $\alpha_c$ 
from the product. Hence, in order to 
evaluate the self-energy of a bound-electron 
quantum cyclotron state to order $\alpha \, \alpha_c^4 m$, 
we need to expand the propagator
$1/(\slashed{\pi} - \slashed{k} - m)$
up to fourth order in $\Gamma_\TT$,
i.e., up to the four-magnetic-vertex term
[term with $n=4$ in Eq.~\eqref{nvertex},
see also Fig.~\ref{fig1}].

A further difficulty arises.
In Fig.~\ref{fig1}, the outer lines 
(outside of the self-energy loop) are 
still fully dressed by the external field;
for hydrogenic bound states, one therefore
uses the known solutions of the Dirac--Coulomb 
equation for the bispinors $| \Psi \rangle$
and $\langle \overline\Psi |$ that enter
the diagram (see, e.g., Chap.~8 of Ref.~\cite{JeAd2022book}).
For the quantum cyclotron problem,
the nonperturbative Dirac solutions
have recently been analyzed in Ref.~\cite{JeAd2022book}.
They read as follows,
\begin{equation}
\label{Psi}
\Psi(\vec r) =
\calN \,
\left( \begin{array}{c}
\psi_\NR(\vec r) \\[2ex]
\dfrac{\vec\sigma \cdot \vec \pi_\TT}{E_D + m} \psi_\NR(\vec r)
\end{array} \right) \equiv
\Psi_{\ell n s}(\vec r) \,.
\end{equation}
where $\psi_\NR$ is the nonrelativistic 
wave functions, and all symbols
will be explained in the following. For the
high-energy part, one notes, in particular,
that the relativistic states are needed for the
untransformed Dirac equation, i.e., in the
form of four-component bispinors. Two-component
solutions obtained after a Foldy-Wouthuysen
transformation~\cite{ZoZhSi2022}
cannot be used as bra and ket vectors
for the fully relativistic self-energy matrix element
in the integrand of Eq.~\eqref{feynman}.
We use the relativistic states
in an approximation where the axial motion
is being neglected, i.e., in the leading 
order in the expansion in powers of 
$\xi_\magnetron$ and $\xi_z$. The Dirac energy is
\begin{equation}
\label{EDD}
E_\DD = m \sqrt{ 1 + \frac{ \omega_c }{m} 
\left( 2 n + s + 1 \right) } \,.
\end{equation}
The nonrelativistic Landau level in the 
symmetric gauge can be separated into a
spinor component $\chi_s$ and 
and a coordinate-space wave function,
\begin{subequations}
\label{wavelns}
\begin{align}
\psi_{\rm NR}(\vec r) =& \;
\psi_{\ell n s}(\vec r) =
\psi_{n \ell}(\vec\rho)  \,
\chi_s \,,
\\[0.1133ex]
\vec\rho =& \; x \, \hate_x + y \, \hate_y \,,
\\[0.1133ex]
\chi_{+1} =& \; \left(
\begin{array}{c} 1 \\ 0
\end{array} \right) \,,
\qquad
\chi_{-1} = \left(
\begin{array}{c} 0 \\ 1
\end{array} \right) \,.
\end{align}
\end{subequations}
The nonrelativistic 
coordinate-space wave function is given as
(see Ref.~\cite{JeAd2022book})
\begin{align}
\label{wavenl}
\psi_{n \ell}(\vec\rho) =& \;
\frac{ 2^{-\tfrac12 (|n-\ell|+1)} }{ \sqrt{\pi} \, \aoc } \,
\sqrt{ \frac{ \min(n, \ell)! }{ \max(n, \ell)! }} \,
\left( \frac{\rho}{\aoc} \right)^{|n - \ell|}
\nonumber\\
& \;
\times
\ii^{| n - \ell |} \,
L^{| n - \ell|}_{\min(n, \ell)}%
\left( \frac12 \, \left( \frac{\rho}{\aoc} \right)^2  \right) \,
\nonumber\\
& \; \times \ee^{\ii (n-\ell) \varphi} \,
\exp\left[ - \frac14 \, \left( \frac{\rho}{\aoc} \right)^2  \right] \,.
\end{align}
Here, $\rho = | \vec \rho |$ and $\varphi = \arctan(y/x)$,
and we use the associated Laguerre polynomials
$L_n^a(x)$~in the conventions of Ref.~\cite{AbSt1972}.
The magnetic Bohr radius is 
\begin{equation}
\aoc = \sqrt{\frac{m c^2}{\hbar \omega_c}} \,
\frac{\hbar}{ m c}
= \frac{\hbar}{\alpha_c m c}
= \sqrt{ \frac{\hbar}{|e| B_\TT}} \,.
\end{equation}
The wave functions $\psi_{\ell n s}(\vec r)$
fulfill the Dirac equation
\begin{subequations}
\begin{align}
H_D \Psi_{\ell n s} =& \; E_D \Psi_{\ell n s}\,,
\\[0.1133ex]
H_D =& \; \vec \alpha \cdot \vec \pi^\parallel_\TT + \beta m \,,
\qquad
\vec \pi^\parallel_\TT = \vec p_\parallel - e \vec A_\TT \,,
\\[0.1133ex]
\vec p_\parallel =& \; p_x \, \hat{\rm e}_x + p_y \, \hat{\rm e}_y \,.
\end{align}
\end{subequations}
Finally, we can give the normalization factor as
\begin{equation}
\calN = \left[ 1 + \frac{2 m E_\NR}{(E_\DD + m)^2} \right]^{-\frac12} \,,
\quad
E_\NR = \frac{\omega_c}{2} \, \left( 2 n + s + 1 \right) \,.
\end{equation}
The relativistic wave function $\Psi$ given in Eq.~\eqref{Psi}
is valid for vanishing axial frequency,
i.e., to leading order in the expansion in the 
ratio $\omega_z/\omega_c$, and can thus 
be used in order to evaluate the high-energy 
part of the quantum cyclotron bound-state self
energy in the leading order in the expansion in powers
of $\omega_z/\omega_c$.

We employ analogous procedures as those that 
were used for the high-energy part of the 
self energy of bound states in hydrogenlike 
systems~\cite{JePa1996}, and 
map the algebra of the quantum cyclotron
states onto a computer algebra system~\cite{Wo1999}.
This enables us to evaluate the matrix 
elements of the vertex terms for the 
high-energy part, where we employ a
noncovariant integration procedure for the 
virtual photon integration contour outlined in 
Sec.~3 of Ref.~\cite{JeSoMo1997}.
The final result for the high-energy part
is (almost) state-independent (except for the
obvious spin-dependence of the leading term) and reads
\begin{equation}
\label{E2HEP}
E^{(2)}_{\rm HEP} = 
\frac{\alpha}{\pi} \, s \, \omega_c + 
\frac{2 \alpha}{3\,\pi} \,
\left[ \ln\biggl(\frac{m}{2\,\epsilon}\biggr) - \frac{13}{72} \right] \,
\frac{\omega_c^2}{m} \,,
\end{equation}
where $\epsilon$ is the (noncovariant) photon 
energy cutoff. The first term in $E^{(2)}_{\rm HEP}$
reproduces the leading anomalous-magnetic-moment 
correction $E^{(0)}_{\rm HEP}$ given in Eq.~\eqref{E0}.

%
%
\subsection{Low--Energy Part}
\label{sec4D}

The appropriate reference state for the
low-energy part is given 
by the nonrelativistic quantum 
cyclotron wave function indicated in Eq.~\eqref{psidef}.
Employing the formalism outlined in Chap.~4 of 
Ref.~\cite{JeAd2022book}, we obtain the expression
\begin{multline}
\label{renlength}
E_{\rm LEP}
=\frac{2\alpha}{3\,\pi} \,
\int_0^\epsilon {\dd} k \, k \,
\\
\times \left< \psi_{k \ell n s} \left| \frac{\pi^i_\PT}{m} \,
\left( \frac{1}{E_0 - H_0 - k} + \frac{1}{k} \right) \,
\frac{\pi^i_\PT}{m} \right| \psi_{k \ell n s} \right> \,,
\end{multline}
where $k = \omega$ is the angular frequency of the 
virtual photon, $\epsilon$ is the photon energy cutoff,
and $| \psi \rangle = | \psi_{k \ell n s} \rangle$
is the reference state.
The sum over $i=1,2,3$ is implied by the Einstein 
summation convention.
We use the relation
\begin{equation}
\label{melem}
\left< \psi \left| \frac{\pi_\PT^i}{m} \,
(H_0 - E_0) \, \frac{\pi_\PT^i}{m} \right| \psi \right> = 
\frac{\omega_c^2}{m} \,,
\end{equation}
which can be shown after expressing the Cartesian components
of the kinetic-momentum operator in terms of 
raising and lowering operators of the cyclotron and magnetron
motions~\cite{BoBrLe1985,BrGa1986,WiMoJe2022,Je2023limit,Je2023mag1}.
Notably, the matrix element given in Eq.~\eqref{melem} 
is state-independent.
After an integration over the photon energy, 
the low-energy part is obtained as
\begin{equation}
\label{ELEP}
E_{\rm LEP}
= \frac{2\,\alpha}{3\,\pi} \,
\frac{\omega_c^2}{m} \,
\ln\left( \frac{\epsilon}{ \alpha_c^2 \, m}\right) 
- \frac{2\,\alpha}{3\,\pi} \, \calM \,,
\end{equation}
where the coefficient of the logarithmic term
contains a logarithmic sum (Bethe logarithm)
over the virtual excitations of the quantum cyclotron state,
\begin{multline}
\calM = 
\left< \phi_r \left| \frac{\pi_\PT^i}{m} \,
(H_0 - E_0) \, 
\ln\left( \frac{\left| H_0 - E_0 \right|}
{\alpha_c^2 \, m} \right) \, \frac{\pi_\PT^i}{m} 
\right| \phi_r \right>  
\\
= \frac{
 \omega_\plus^3 \, \ln\left( \frac{\omega_\plus}{\omega_c} \right)
- \omega_\minus^3 \, \ln\left( \frac{\omega_\minus}{\omega_c} \right) }%
{m \, (\omega_\plus - \omega_\minus)} 
+ \frac{\omega_z^2}{2 m} \, \ln\left( \frac{\omega_z}{\omega_c} \right) \,.
\end{multline}
In the simplification of the expressions, 
the following identities prove to be extremely useful,
\begin{align}
\omega_\plus - \omega_\minus = & \;
\sqrt{ \omega_c^2 - 2 \omega_z^2 } \,,
\\
2 \, \omega_\plus \, \omega_\minus = & \; \omega_z^2 \,.
\end{align}
Furthermore, it is very interesting to observe that,
in the limit $\omega_z \to 0$, which 
implies $\omega_\plus \to \omega_c$ and $\omega_\minus \to \omega_c$,
the Bethe-logarithm matrix element $\calM$ vanishes.
The first nonvanishing contribution to $\calM$
appears at order~$\xi_z^4$.

%
%
\subsection{Self--Energy Shift}
\label{sec4E}

After adding the high- and low-energy contributions,
the dependence on the photon energy cutoff $\epsilon$
cancels [see Eqs.~\eqref{E1HEP},~\eqref{E2HEP} and~\eqref{ELEP}]. 
The total self-energy shift $E_{\rm SE}$,
up to order $\alpha \, \alpha_c^4 m$, is 
obtained as follows,
\begin{equation}
\label{ESE}
E_{\rm SE}  =
E^{[1]}_{\rm HEP} + E^{[2]}_{\rm HEP} + E_{\rm LEP} 
= \sum_{i=1}^6 \calT_i \,,
\end{equation}
where the six individual contributions (together with 
their respective expansion in powers of $\xi_z$) are
\begin{subequations}
\begin{align}
\calT_1 = & \;
\frac{\alpha}{\pi} \, s \, \omega_c = 
\frac{\alpha}{\pi} \, \alpha_c^2 m \, s \,,
\\
\calT_2 = & \;
\frac{\alpha}{\pi} \, \alpha_c^4 m \,
\left[ \frac23 \ln\left(\alpha_c^{-2} \right) - 
\frac23 \, \ln(2) - \frac{13}{108} \right] \,,
\\
\calT_3 = & \;
- \frac{\alpha}{8 \pi} \frac{s \omega_c \omega_z}{m} (k + \tfrac12) 
= -\frac{\alpha}{8 \pi} \alpha_c^4 m 
s \, \xi_z^2 \, (k + \tfrac12) \,,
\\
\calT_4 = & \;
- \frac{\alpha}{4 \pi} \frac{\omega_z^2 s}{m}
\frac{ \omega_\plus (n + \tfrac12) + \omega_\minus (\ell + \tfrac12) }%
{\omega_\plus - \omega_\minus} 
\nonumber\\
= & \; \frac{\alpha}{\pi} \alpha_c^4 m \,
\left[ -\frac18 (2n+1) s \, \xi_z^4 \right] + \calO(\xi_z^6) \,,
\\
\calT_5 = & \; - \frac{\alpha}{3\,\pi} \,
\frac{\omega_z^2}{m} \, \ln\left( \frac{\omega_z}{\omega_c} \right) 
\nonumber\\
= & \;  \frac{\alpha}{\pi} \alpha_c^4 m \,
\left[ -\frac23 \, \xi_z^4 \, \ln(\xi_z) \right] \,,
\\
\calT_6 = & \;
- \frac{2\,\alpha}{3\,\pi} \,
\frac{ \omega_\plus^3 \, \ln\left( \frac{\omega_\plus}{\omega_c} \right)
- \omega_\minus^3 \, \ln\left( \frac{\omega_\minus}{\omega_c} \right)
}{ m (\omega_\plus - \omega_\minus) } 
\nonumber\\
=& \; \frac{\alpha}{\pi} \alpha_c^4 m \frac{\xi_z^4}{3} + \calO(\xi_z^6) \,.
\end{align}
\end{subequations}
The leading (state-independent)
logarithmic contribution to the Lamb shift 
of a quantum cyclotron state is 
\begin{equation}
E_L = 
\frac{2\,\alpha}{3\,\pi} \, m \, \frac{\omega_c^2}{m^2} \,
\ln\left(\alpha_c^{-2} \right) 
= \frac{2\,\alpha}{3\,\pi} \, \alpha_c^4 \, m \,
\ln\left(\alpha_c^{-2} \right) \,.
\end{equation}
It is reassuring to see that the only state-dependent 
contributions to the QED energy shift of order
$\alpha \, \alpha_c^4 \, m$ come from the anomalous magnetic 
moment.

The final results of our investigations 
can be summarized in the following,
concise form, encapsulating the leading 
coefficients in the self energy shift 
given in Eq.~\eqref{ESEcoeff},
\begin{subequations}
\label{ESEres}
\begin{equation}
\label{E4}
E_{\rm SE} = 
\frac{\alpha}{\pi} \, \alpha_c^2 m \calA_{20} + 
\frac{\alpha}{\pi} \, \alpha_c^4 m 
\left[ \calA_{41} \, \ln( \alpha_c^{-2} ) 
+ \calA_{40} \right] \,,
\end{equation}
where the coefficients are, except for $\calA_{20}$,
state-independent, and read as follows
in the leading order of the expansion
in powers of $\xi_z$,
\begin{align}
\label{E4b}
\calA_{20} =& \; s \,, \qquad
\calA_{41} = \frac23 \,, \\
\label{E4c}
\calA_{40} =& \; -\frac23 \ln(2) -\frac{13}{108} 
+ \calO(\xi_z^2) \,,
\end{align}
\end{subequations}
We also evaluate partial results for the 
dependence of the $\calA_{40}$ coefficient
on the axial frequency. These results
are partial, because the treatment of the 
high-energy part of the self energy employed by us 
is valid only to leading order in $\xi_z$.
The corrections evaluated by us add up to the 
partial higher-order (h.o.) result
\begin{multline}
\label{E4ho}
\left. \calA_{40} \right|_{\mbox{h.o.}}
= \left. \calA_{40} \right|_{\xi_z=0} 
- \frac18 \, \left( k + \frac12 \right) \, s \, \xi_z^2
\\
+ \left[ \frac13 - \frac18 (2n+1) s - \frac23  \ln(\xi_z) \right] \, \xi_z^4
+ \calO(\xi_z^6) \,.
\end{multline}

\begin{figure}[t!]
\begin{center}
\begin{minipage}{1.0\linewidth}
\begin{center}
\includegraphics[width=0.87\linewidth]{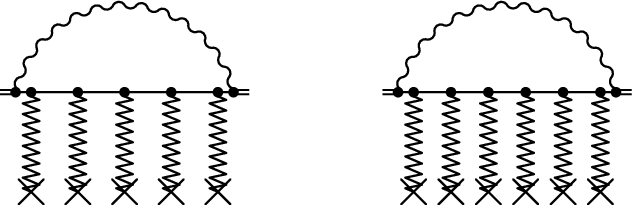}
\caption{\label{fig2}
The diagrams with five magnetic vertices (left),
and  six magnetic vertices (right) contribute to the 
state-dependent, logarithmic term of 
order $\alpha \, \alpha_c^6 m \, \ln(\alpha^{-2})$,
as discussed in Eq.~\eqref{A61}.}
\end{center}
\end{minipage}
\end{center}
\end{figure}

%
%
\subsection{Higher--Order Logarithmic Term}

It is somewhat surprising to see that the 
coefficients $\calA_{41}$ and $\calA_{40}$ are
state-independent in the leading order 
in the expansion in $\xi_z$. 
Because the axial frequency is small compared to 
the cyclotron frequency, this observation
raises the question at which order in the 
expansion in $\alpha_c$ (i.e., in the main 
cyclotron frequency expansion parameter)
any state dependence is actually incurred.
With some effort, one
can obtain the leading logarithmic 
terms in the sixth order in $\alpha_c$ 
from the six-vertex correction (see Fig.~\ref{fig2}).
We obtain, after algebraic simplification, 
the result
\begin{equation}
\label{A61}
\delta E_{\rm SE} = \frac{\alpha}{\pi} \, \alpha_c^6 m \,
\calA_{61} \, \ln( \alpha_c^{-2} ) \,,
\quad
\calA_{61} =
2 n + 1 - \frac{4s}{3} \,.
\end{equation}
This result depends on the spin orientation 
of the reference state, just like $\calA_{20}$,
and also grows with the principal quantum number $n$,
which is the quantum number that counts the 
cyclotron excitations.
Further details of the derivation will
be presented elsewhere~\cite{Je2023prep}.

For hydrogenic bound states, the higher-order 
coefficients typically decrease with the 
principal quantum number~\cite{LBEtAl2003};
for quantum cyclotron states, the dependence is 
reversed. The physical reason for this is that in
hydrogen, higher excited states have a lesser
expectation values of the momentum square, 
and are, in that sense, less relativistic and
subjected to a lesser extent to relativistic 
and quantum electrodynamic corrections.
Specifically, in a hydrogenic state with 
principal quantum number $n$, the typical 
momentum scale is $Z\alpha m/n$, where 
$Z$ is the nuclear charge number, $\alpha$ is the 
fine-structure constant, and $m$ is the electron mass.
For a quantum cyclotron state, the momentum 
scale is $\alpha_c \, m \, \sqrt{n}$, 
where $\alpha_c$ is defined in Eq.~\eqref{defalpha}.
So, it is natural that $\calA_{61}$ increases with the 
quantum cyclotron quantum number $n$.

%
%
\section{Conclusions}
\label{sec5}

In this paper, we have discussed the QED energy shifts
of quantum cyclotron levels. 
We start from a very concise recap of the 
main ingredients of quantum cyclotron levels in 
Sec.~\ref{sec2}, with vacuum-polarization
effects discussed in Sec.~\ref{sec3} and 
the dominant self-energy shift discussed in Sec.~\ref{sec4}.
In the Penning trap, the 
rotational symmetry of the hydrogen and 
atomic bound-state problem is lost, and
only the axial symmetry of the magnetic trap field
remains.
Hence, one formulates the bound states
using spin-up and spin-down fundamental spinors
[see Eq.~\eqref{psidef}], 
rather than the spin-angular functions known from
atomic bound-state theory (see Chap.~6 of Ref.~\cite{JeAd2022book}).

The kinetic momentum operator 
$\vec\pi_\PT$ given in Eq.~\eqref{piPTdef} 
can easily be decomposed
into raising and lowering operators
for the cyclotron, axial, and magnetron motions.
Hence, one can express the matrix elements of the 
radiatively corrected relativistic Hamiltonian
given in Eq.~\eqref{Hkappa} in terms
of the quantum numbers $k$, $\ell$, $n$ and $s$
(see also Refs.~\cite{BoBrLe1985,BrGa1986,WiMoJe2022,Je2023limit,Je2023mag1}).
One adds the high-energy contribution
due to the anomalous magnetic moment
from Eq.~\eqref{E1HEP},
and the high-energy contribution from the 
terms with up to four magnetic vertices,
as given in Eq.~\eqref{E2HEP},
to the low-energy term listed in Eq.~\eqref{ELEP}.
The complete
self-energy shift of order $\alpha \, \alpha_c^4 \, m$
is given in Eq.~\eqref{ESE}.
By considering diagrams with 
up to six magnetic vertices (see Fig.~\ref{fig2}),
as a significant further result, one obtains a state-dependent,
higher-order logarithmic binding correction 
of order $\alpha \, \alpha_c^6 \, m \ln(\alpha_c^{-2})$
in Eq.~\eqref{A61}.

A few words on the experimental and phenomenological 
relevance of the higher-order binding corrections
calculated here are in order.
Because of the scaling with higher powers 
of the coupling parameter $\alpha_c$, 
the effects become more pronounced
in stronger magnetic fields.
In current Penning trap experiments~\cite{FaEtAl2023},
field strengths of the order of $B_\TT \approx 5.3 \, {\rm T}$ 
are employed, resulting in
$\omega_c \approx 2 \pi \times 148 \, {\rm GHz}$
and cyclotron coupling parameter 
$\alpha_c \approx 3.5 \times 10^{-5}$,
which implies $\ln[\alpha_c^{-2}] \approx 20.5$.
With an axial frequency of the order of
$\omega_z \approx 2 \pi \times 114 \, {\rm MHz}$,
one has $\xi_z \approx 0.028$.

The higher-order one-loop 
binding corrections to the quantum cyclotron energy 
levels calculated here scale as follows.
We have in the fourth order in $\alpha_c$,
from Eq.~\eqref{E4},
\begin{equation}
\delta E^{(4)} =  
\frac{\alpha}{\pi} \, \alpha_c^4 m
\left[ \calA_{41} \, \ln( \alpha_c^{-2})
+ \calA_{40} \right] \,,
\end{equation}
where the coefficients $\calA_{41}$ and $\calA_{40}$
are state-independent in the 
leading order in the expansion in $\xi_z$
[see Eqs.~\eqref{E4b} and~\eqref{E4c}].
Quantum cyclotron levels are displaced from each 
other by an energy $\omega_c = \alpha_c^2 m$.
Hence, the relative shift of the 
cyclotron frequency due to the quantum electrodynamic
effects is 
\begin{equation}
\chi^{(4)} \sim \frac{\delta E^{(4)}}{\alpha_c^2 \,m} \,,
\qquad
\chi^{(4)} = \frac{\alpha}{\pi} \, \alpha_c^2 
\ln( \alpha_c^{-2}) \,.
\end{equation}
The nonlogarithmic coefficient $\calA_{40}$ receives
corrections of order $\xi_z^2$ according to 
Eq.~\eqref{E4ho}. Parametrically, these additional
terms lead to a relative energy shift of the 
order of $\chi^{(z)}$, where
\begin{equation}
\chi^{(z)} = 
\frac{\alpha}{\pi} \, \alpha_c^2 \, \xi_z^2 
\end{equation}
for quantum cyclotron levels.
Finally, the higher-order binding corrections
given in Eq.~(61) gives rise to a relative energy shift
described by the parameter $\chi^{(6)}$, where 
\begin{equation}
\chi^{(6)} = 
\frac{\alpha}{\pi} \, \alpha_c^4 \, 
\ln( \alpha_c^{-2}) \,.
\end{equation}
For $B_\TT \approx 5.3 \, {\rm T}$ and
$\omega_z \approx 2 \pi \times 114 \, {\rm MHz}$,
i.e., the parameters of Ref.~\cite{FaEtAl2023},
one has
\begin{align}
\left. \chi^{(4)} \right|_{B_T = 5.3 \, {\rm T}} = & \;
5.8 \times 10^{-11} \,,
\\
\left. \chi^{(z)} \right|_{B_T = 5.3 \, {\rm T}} = & \;
2.1 \times 10^{-15} \,,
\\
\left. \chi^{(6)} \right|_{B_T = 5.3 \, {\rm T}} = & \;
6.9  \times 10^{-20} \,.
\end{align}
In view of these results, we can say that 
the absence of a state-dependence of $\calA_{41}$ 
and $\calA_{40}$ (in the leading order in $\xi_z$) 
is crucial for the validity of 
the evaluation of the recent experiment~\cite{FaEtAl2023},
as any dependence on $n$ could have 
easily shifted the determination of the 
cyclotron frequency (and of the electron $g$ factor)
on the level of $10^{-11}$, which is larger than the 
experimental uncertainty reported in Ref.~\cite{FaEtAl2023}
by roughly {\em two} orders of magnitude. 
The absence of a state-dependence of $\calA_{41}$
and $\calA_{40}$, in the leading order in $\xi_z$,
is one of the most important results of the current
investigation.

The corrections parameterized by $\chi^{(z)}$
and $\chi^{(6)}$ are not relevant at current 
experimental conditions~\cite{FaEtAl2023}.
However, according to Table~1 of Ref.~\cite{BaEtAl2018mag},
it is clear that magnetic field strengths in
excess of $30 \, {\rm T}$ are current maintained 
in continuous (DC) mode by a number of laboratories
around the World.
One of the most impressive results available to date
is the $45.5 \, {\rm T}$ field reported in 
Ref.~\cite{HaEtAl2019mag}.
It is thus instructive to carry out calculations
for a magnetic field of $B_\TT = 30 \, {\rm T}$,
with the results
\begin{align}
\left. \chi^{(4)} \right|_{B_T = 30 \, {\rm T}} = & \;
3.0 \times 10^{-10} \,,
\\
\left. \chi^{(z)} \right|_{B_T = 30 \, {\rm T}} = & \;
2.1 \times 10^{-15} \,,
\\
\left. \chi^{(6)} \right|_{B_T = 30 \, {\rm T}} = & \;
2.0  \times 10^{-18} \,.
\end{align}
For these conditions, the correction of
order $\alpha \, \alpha_c^6 \, \ln[\alpha_c^{-2}]$ 
could become relevant, when experimental techniques 
are combined with 
modern spectroscopic techniques~\cite{PrEtAl2013}.
It is also very important to realize that 
state-dependent coefficients grow linearly
with the cyclotron quantum number $n$,
and axial quantum number $k$
[see Eqs.~\eqref{E4ho} and~\eqref{A61}].
The corrections thus become much more 
important for higher excited cyclotron states.
We also observe that the mass $m$ of the trapped 
particle cancels out in the relative corrections
denoted by the symbols $\chi^{(4)}$,
$\chi^{(z)}$, and $\chi^{(6)}$, discussed above;
in other words, the quantities $\chi^{(4)}$,
$\chi^{(z)}$, and $\chi^{(6)}$
are functions of the coupling parameter
$\alpha_c$ only. For given magnetic field,
the coupling parameter $\alpha_c$
is inversely proportional to the 
trapped particle mass $m$ [see Eq.~\eqref{defalpha}],
in view of the relation $\alpha_c = \sqrt{|e| B_\PT}/m$. 
Hence, for hydrogen-like and lithium-like
bound systems (ions) in a Penning trap,
the quantum electrodynamic effects 
scale according to the dependence of $\alpha_c$ on 
the mass of the trapped ion.

Three final remarks are in order.
{\em (i)} First, we reemphasize that vacuum-polarization contributions
can be safely neglected, as already discussed 
near the beginning of Sec.~\ref{sec3}. {\em (ii)}
Second, we would like to 
remind the reader that modifications of the QED
shifts due to the 
cylinder walls of the Penning trap~\cite{BoBrLe1985,Je2023limit}
have not been considered in the current work.
We here work with the full photon propagator that is 
unperturbed by the external conditions due to the 
cylinder walls of the Penning trap.
Because the average spatial extent of a quantum cyclotron
state is only a tiny fraction of the trap dimension,
this approximation is well justified,
with limitations being discussed 
in Refs.~\cite{BoBrLe1985,Je2023limit}.
{\em (iii)}
Relativistic Bethe logarithm corrections to the 
leading one-loop terms are of order 
self-energy shift of order $\alpha \, \alpha_c^6 \, m$
while the correction obtained
in Eq.~\eqref{A61} is enhanced by the logarithm $\ln(\alpha_c^{-2})$.
The evaluation of relativistic Bethe logarithms,
for quantum cyclotron states
complementing work on hydrogenic levels~\cite{Pa1993,JePa1996},
would be an inspiration for future studies~\cite{Je2023prep}.

%
%
\section*{Acknowledgments}

The authors acknowledge helpful conversations with 
Professors Gerald Gabrielse and Krysztof Pachucki,
and support from the
Templeton Foundation (Fundamental Physics Black Grant,
Subaward 60049570 of Grant ID \#{}61039),
and from the National Science Foundation (grant PHY--2110294).

\end{document}